\documentclass[twocolumn,10pt,showpacs,preprintnumbers,amsmath,amssymb]{revtex4}
\usepackage{graphicx}% Include figure files
\usepackage{dcolumn}% Align table columns on decimal point
\usepackage{bm}% bold math
\usepackage{color}
\usepackage{placeins}
\usepackage{lineno,hyperref}
\usepackage{multirow}
\usepackage{amsmath}
\begin{document}

% The following information is for internal review, please remove them for submission

% the following line is for submission, including submission to the arXiv!!
%\hspace{5.2in} \mbox{Fermilab-Pub-04/xxx-E}

%\title{The role of chromo-electromagnetic field fluctuations on $J/\psi$ production in Pb-Pb collisions at $2.76$ TeV at the LHC} 
\title{Heavy flavour Langevin diffusion with the chromo-electromagnetic field fluctuations in the quark-gluon plasma} 
%\date{\today}

\author{\it Ashik Ikbal Sheikh}
\email{ashikhep@gmail.com}
\address {Variable Energy Cyclotron Centre, HBNI, 1/AF Bidhan Nagar, Kolkata - 700064, India}
%\address {Homi Bhabha National Institute, Training School Complex, Anushaktinagar, Mumbai - 400085, India}
\author{\it Zubayer Ahammed}
\email{za@vecc.gov.in}
\address {Variable Energy Cyclotron Centre, HBNI, 1/AF Bidhan Nagar, Kolkata - 700064, India}
%\address {Homi Bhabha National Institute, Training School Complex, Anushaktinagar, Mumbai - 400085, India}

\begin{abstract}
The chromo-electromagnetic field is produced due to the motion of partons in a quark-gluon plasma created by relativistic heavy-ion collisions. The fluctuations in the produced chromo-electromagnetic field are important, since they cause heavy quarks to gain energy in the low velocity limit. We study the effect of such fluctuations on heavy quark diffusion in quark-gluon plasma within the framework of Langevin dynamics under the background matter described by the ($3+1$)-dimensional relativistic viscous hydrodynamics. Theoretical calculations of the nuclear modification factor ($R_{AA}$) of heavy mesons ($D$ and $B$ mesons), with the effect of these fluctuations, are compared with experimental measurements in $Au-Au$ collisions at $\sqrt{s_{NN}} = 200$ GeV by the STAR experiment at the BNL Relativistic Heavy Ion Collider (RHIC) and $Pb-Pb$ collisions at $\sqrt{s_{NN}} = 2.76$ and $5.02$ TeV by the ALICE and CMS experiments at the CERN Large Hadron Collider (LHC). We find a significant effect of these fluctuations in describing the the measured $R_{AA}$ of $D$ and $B$ mesons in both RHIC and LHC energies.

\end{abstract}

\maketitle

\section{Introduction}
\label{intro}
The ongoing relativistic heavy-ion programs at the Relativistic Heavy Ion Collider (RHIC) at BNL and at the Large Hadron Collider (LHC) at CERN are aimed to create a new novel state of matter where the bulk properties are governed by light quarks and gluons degrees of freedom~\cite{Shuryak,Muller} and this new phase of matter is well known as quark-gluon plasma (QGP). The space-time evolution of the QGP is successfully described by relativistic hydrodynamic calculations~\cite{Gyulassy,Teaney,Huovinen,Hirano,Song}.
There are several probes to extract informations about the properties of the QGP. The collective properties of the QGP are obtained by studying the soft probes like light hadron distributions at low transverse momentum, whereas the microscopic properties are obtained by studying the hard probes like heavy quarks, jets and quarkonia~\cite{Vitev}.

Here, we focus on heavy quarks (charm and bottom), which are mainly produced in the primordial hard scatterings in the relativistic heavy-ion collisions. The heavy quarks do not constitute the bulk of the QGP medium and hence they act as impurities in the medium. As they are produced early, they encounter the full space-time evolution of the medium and as a result of that they might preserve a memory of their interaction history which makes them a crucial probe of the QGP medium.  The thermalization time scale for heavy quarks is larger than that of light quarks by a factor of $M_{Q}/T \sim 5-20$~\cite{RappBook,Moore,Hees,Cao}, where $M_{Q}$ is the mass of the heavy quarks and $T$ being the typical temperature of the QGP medium formed at RHIC and LHC. Since $M_{Q}$ is much larger than $T$ and as well as the constituent masses of the QGP medium, relaxation time of heavy quarks is also larger
than that for light quarks which justifies a soft-scattering approximation. Under such scenario, heavy quarks execute Brownian motion inside the QGP medium and their motions are well treated in the framework of Fokker-Plank equation~\cite{RappBook,Moore,Svetitsky,MGM1}. The Fokker-Planck equation is understood by stochastic Langevin processes~\cite{RappBook,Dunkel,MinHe} which makes a great simplification of in-medium heavy-flavour (HF) dynamics. Such formalism is very extensively employed to calculate the experimentally measured nuclear modification factor ($R_{AA}$) of HF mesons ($D$ and $B$) in the literature~\cite{Moore,Hees,Cao,MGM1,Adronic,Prino,Rapp,Cao:2018ews,Dong:2019unq,Santosh14,Li18,LiWang,Hees2,Gossiaux,GossiauxPoS,Akamatsu,Alberico,Young,MinHe2,Lang,Cao2,Cao3,Xu,MinHe3,Santosh,Santosh17,Xu18}. 

During the process of diffusion, the heavy quarks lose energy via ellastic collisions with the medium partons~\cite{TG,BT,Alex,PP} and bremsstrahlung gluon radiations~\cite{dokshit01,dead,abir12,Fochler:2008ts,Gossiaux10,wicks07,W.C,Vitev,AJMS}. These two effects are translated into the behaviour of HF meson spectra as measured in the RHIC~\cite{Rhic} and LHC~\cite{ALICE_D,ALICE_DNpart,CMS_D_2TeV,CMS_D_5TeV,CMS_B_5TeV} experiments. The estimations of heavy quark energy losses in the QGP were obtained by considering the QGP medium in an average fashion, i.e., microscopic fluctuations were ignored. The QGP, being a statistical ensemble of ambulant light quarks and gluons, is described by widespread stochastic chromo-electromagnetic field fluctuations. Such fluctuations couple to external disturbances, and the response of the medium is greatly affected due to these disturbances. The field fuctuations in a nonrelativistic classical plasma causes an energy gain of a charged particle passing through the plasma and it has been worked out by several authors~\cite{Gasirowicz,Sitenko,Akhiezer,Kalman,Thompson,Ichimaru}. The parton energy loss in the QGP by considering the effect of stimulated gluon emission and thermal absorption has been reported in Ref.~\cite{wangwang}. When an energetic heavy quark passes through the QGP medium, it encounters the fluctuations of chromo-electromagnetic field generated by the moving partons in the medium which causes the heavy quark to gain energy and the gained energy is substantial in lower momentum region~\cite{Fl}. The effect of these chromo-electromagnetic field fluctuations is extremely important to study the heavy quark propagation in QGP and these fluctuations have a great impact on the experimentally measured HF suppressions as shown in our previous work where we considered simple 1D expansion for the QGP medium evolution and the detailed dynamics of the heavy quarks in the medium were not considered~\cite{Ours}.

In this article, for the first time, we investigate the effect of these field fluctuations on heavy quarks in the framework of Langevin simulation. We solve numerically the Langevin equation for heavy quark in the evolving hydrodynamic background. For the modelling of hydrodynamic background, we use ($3+1$)-dimensional relativistic viscous hydrodynamics code, vHLLE~\cite{Karpenko}. We calculate nuclear modification factor, $R_{AA}$ of HF mesons and compared with experimental measurements in $Au-Au$ collisions at $\sqrt{s_{NN}} = 200$ GeV by the STAR experiment and $Pb-Pb$ collisions at $\sqrt{s_{NN}} = 2.76$ and $5.02$ TeV by the ALICE and CMS experiments.

The article is organized as follows: In the next section, we discuss the ingredients for Langevin simulation of heavy quark diffusion in hydrodynamic background, i.e., the transport coefficients, initial distribution and fragmentation of heavy quarks, the chromo-electromagnetic field fluctuations and the description of hydrodynamic evolution. In Sec.\ref{sec3}, we discuss our results. Sec.\ref{sec4} is devoted to summary and conclusions.

\section{Langevin Simulation for Heavy Flavour Diffusion}
\label{sec2}
In this section, we start with a brief description of Langevin dynamics (\ref{sec2.1}), the transport coefficients (\ref{sec2.2}), the effect of chromo-electromagnetic field fluctuations on heavy quarks (\ref{sec2.3}) and initial heavy quark production and fragmentation (\ref{sec2.4}). In Sec.\ref{sec2.5}, we introduce the model for hydrodynamic evolution of background QGP medium.

\subsection{Langevin dynamics}
\label{sec2.1}
Consider a heavy quark propagates in a QGP medium of light partons at a temperature, $T < M_{Q}$, its momentum change due to collisions with the medium constituents is relatively small. Under this soft-scattering approximation, the Boltzmann equation describing the motion of the heavy quark in the QGP medium reduces to
the Fokker-Planck equation. The Fokker-Planck equation is solved by stochastic Langevin equation~\cite{RappBook,Moore,Cao,Svetitsky}:

\begin{equation}
\label{xupdate}
dx_{i}  =  \frac{p_{i}}{E}dt
\end{equation}

\begin{equation}
\label{pupdate}
dp_{i}  =  -\gamma p_{i}dt + \rho_{i} \sqrt {2Ddt}
\end{equation}
where $dx_{i}$ and $dp_{i}$ refer to the updates of the position and momentum of the heavy quark in each time step $dt$ with $i = x, y$ and $z$, denotes the three components of position in Cartesian coordinates. We have assumed here a diagonal form for the diffusion matrix as used in the past by several authors~\cite{Hees,Cao,Santosh17}. $\rho_{i}$ is the standard Gaussian noise variable which is distributed randomly according to,

\begin{equation}
w(\rho) =  \frac{1}{(2\pi)^{3/2}}\exp(-\rho^{2}/2)
\end{equation}
The random variable $\rho_{i}$ satisfies the relation, $<\rho_{i} > = 0$ and $ <\rho_{i}\rho_{j}> = \delta(t_{i}-t_{j}) $.
In the following, we use the post-point discretization scheme where the equilibrium condition takes the form of fluctuation-dissipation theorem $D = \gamma E T$, where $\gamma$ and $D$ are the drag and diffusion coefficient respectively, that govern the interaction between the heavy quark and the medium, $E = \sqrt {p^2+M_{Q}^2}$ is the energy of the heavy quark. We have checked that in long time limit, the heavy quark phase space distribution function converges to the equilibrium Boltzmann-Juttner
function $e^{-E/T}$. The Langevin equations, Eq.\ref{xupdate} and Eq.\ref{pupdate}, are valid for a static
medium. In our context, for the evolving medium we perform a Lorentz boost to each heavy quark into the local rest frame of the fluid cell through which
it propagates and the position and momentum
are updated according to Eq.\ref{xupdate} and Eq.\ref{pupdate}. After that, we boosted back to the laboratory rest frame to obtain the heavy quark phase space coordinates. In our simulation, we stop the Langevin evolution when the temperature of the background medium drops to $150$ MeV, where particle spectra are calculated in statistical emission model~\cite{Chojnacki}.

\subsection{Transport coefficients}
\label{sec2.2}
The transport coefficients are important quantities containing the dynamics of the heavy quarks propagating through the QGP medium. The estimation for these transport coefficients of heavy quarks in the QGP medium is not yet a settled issue~\cite{Moore,Santosh17,Xu18}. The drag coefficient is related to energy loss, $-dE/dx$ of the heavy quark propagating in the QGP medium. We have estimated the drag coefficient, $\gamma$, for heavy quark by using $\gamma = \frac{1}{p}(-\frac{dE}{dx})$~\cite{MGM05,Santosh10}. Here we have considered collisional as well as radiative energy losses. The heavy quarks collide elastically with the thermal partons (light quarks and gluons) in the QGP medium. We have used the most detailed calculation for these collisional processes as derived by Peigne and Pashier~\cite{PP}. Besides the collisional energy loss, the heavy quarks also suffer radiative energy loss which is the dominant process for a fast parton moving inside the QGP medium. The heavy quark energy loss gets suppressed compared to the case of the light quarks due to dead cone effect. For heavy quark radiative energy loss, we use the calculation as discussed in Ref.~\cite{AJMS}, which is derived based on generalized dead cone approach and the gluon emission probability.

\begin{figure} % figuur 1
	\begin{minipage}{\columnwidth}
		\centering
		\includegraphics[width=\textwidth]{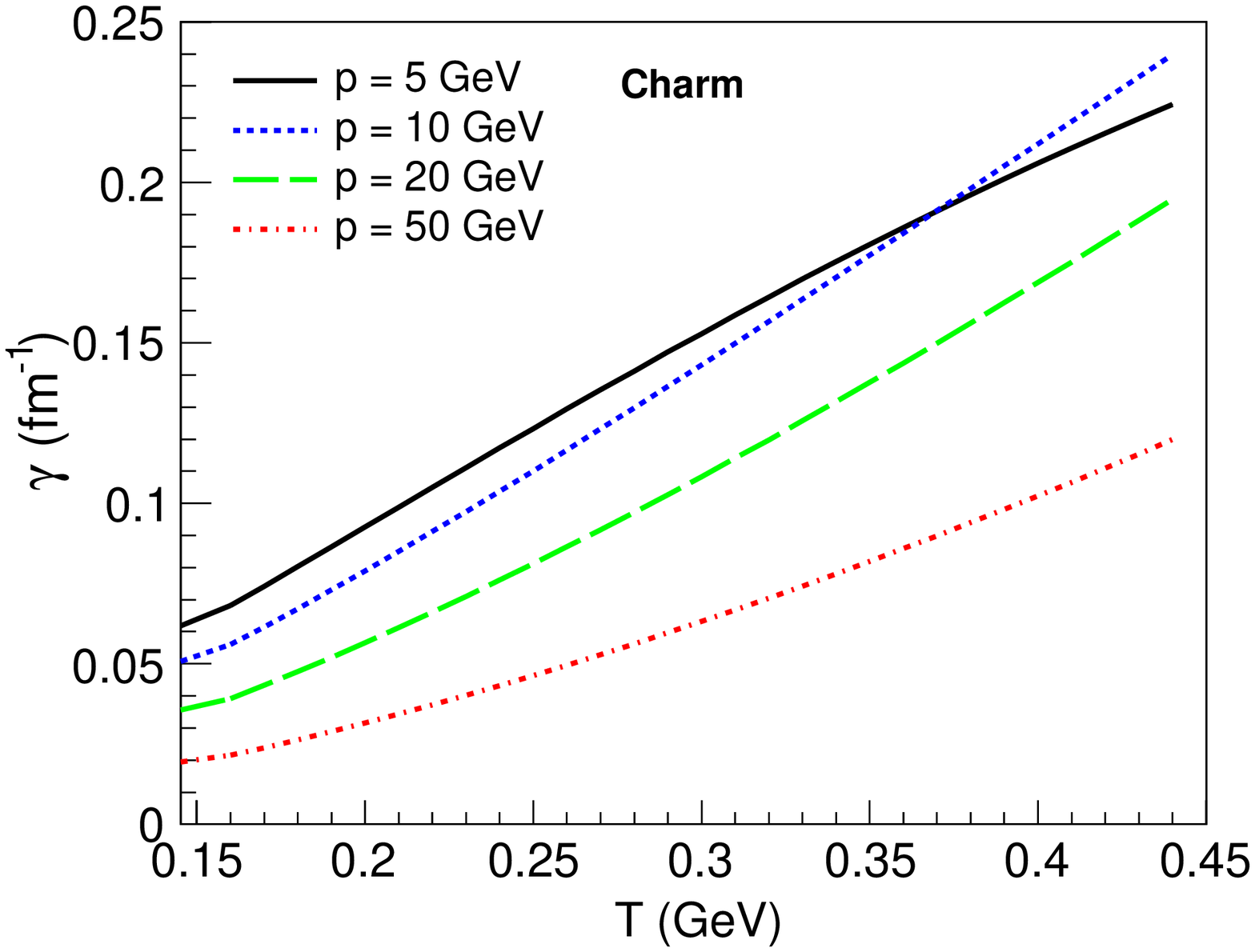}
		\caption{Drag coefficient of a charm quark with momentum $p = 5, 10, 20$ and $50$ GeV as a function of $T$.}
		\label{dragc}
		\includegraphics[width=\textwidth]{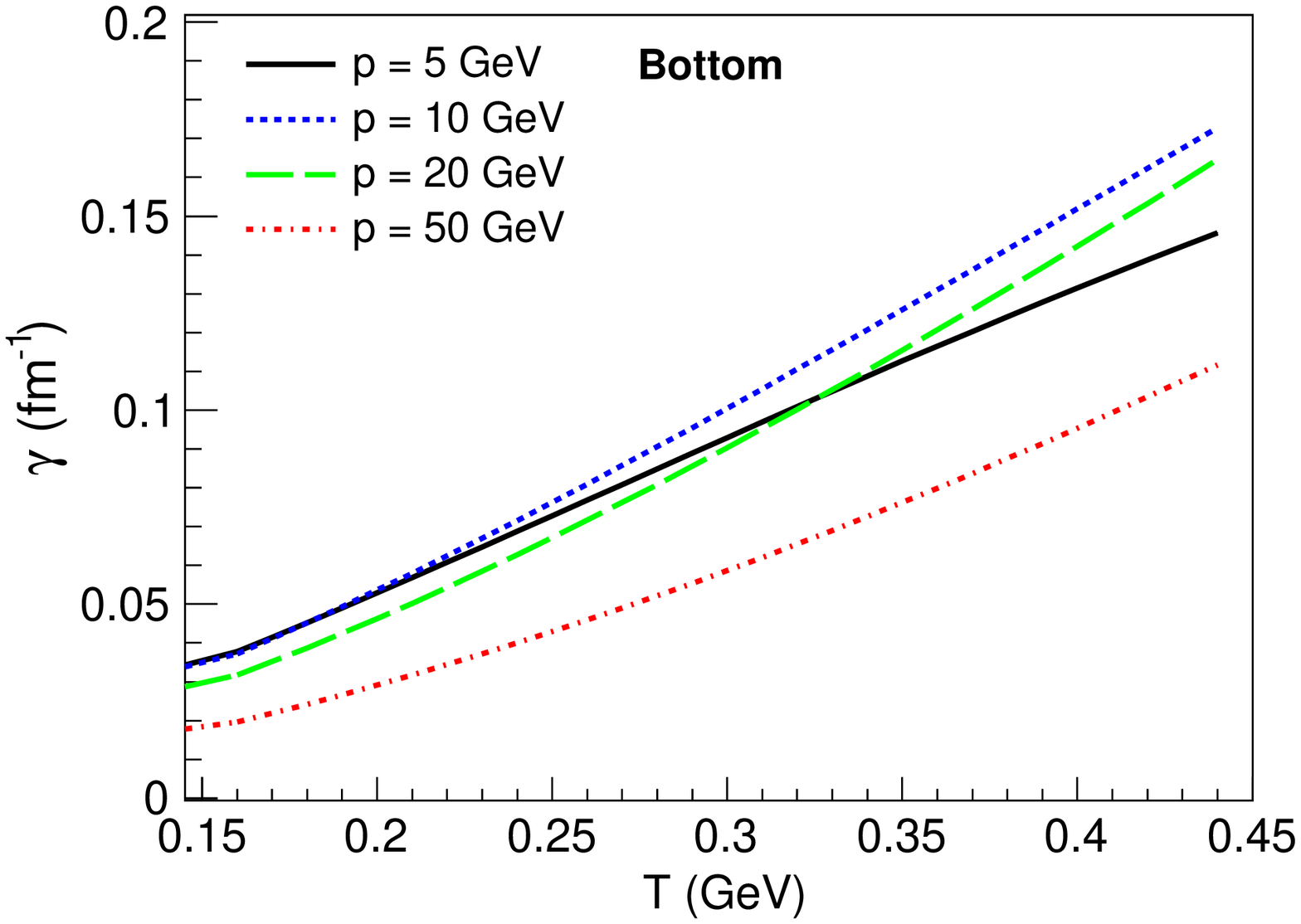}
		\caption{Drag coefficient of a bottom quark with momentum $p = 5, 10, 20$ and $50$ GeV as a function of $T$.}
		\label{dragb}
	\end{minipage}	
\end{figure}

In Figs.\ref{dragc} and \ref{dragb}, we have shown the $T$ dependence of the drag coefficient for a charm and bottom quark of momentum $p = 5, 10, 20$ and $50$ GeV respectively. We use charm quark mass $M_{c} =
1.25$ GeV and bottom quark mass $M_{b} = 4.2$ GeV in our calculations. The values of the drag coefficients, both for charm and bottom quark, increase as temperature of the medium increases. In other words, the heavy quarks get dragged more inside the medium when temperature of the medium is higher. It is also observed that the heavy quarks with large momentum experience less dragging force due to the medium compared to the low momentum heavy quarks.

\subsection{Chromo-electromagnetic field fluctuations}
\label{sec2.3}
The energy loss of a parton propagating in QGP medium is obtained through the work done by the retarding
forces acting on the parton in the plasma due to the chromo electromagnetic field generated by the parton itself because of its motion. Since QGP is a statistical system of mobile light quarks and gluons, it is characterized by widespread stochastic chromo-electromagnetic field fluctuations. Nevertheless, the calculations for heavy quark energy loss (collisional and radiative) do not take into account these chromo-electromagnetic field fluctuations. The fluctuations of the chromo-electromagnetic field causes a statistical change in the energy of the moving heavy quark inside plasma and the velocity of the heavy quark under the influence of this field. As a consequence of that, the heavy quark gains energy and the leading log (LL) contribution of this gained energy is obtained by using semiclassical approximation as~\cite{Fl}:

\begin{equation} 
\left(\frac{dE}{dx}\right)_{\mbox{fl}}^{\mbox{LL}} = 2\pi C_F\alpha_{s}^{2}\left(1+\frac{n_f}{6}\right)\frac{T^3}{Ev^2}\ln{\frac{1+v}{1-v}} \ln{\frac{k_{\mbox{max}}}{k_{\mbox{min}}}},
\end{equation}
where $k_{\mbox{max}} = \mbox{min}\left[E, {2q(E+p)}/{\sqrt{M^2+2q(E+p)}} \right]$ with
$q \sim T$ is the typical momentum of the thermal partons (light quarks and gluons) in the QGP and $k_{\mbox{min}} = \mu_g $ is the Debye mass. We have considered the effect these fluctuations while solving the Langevin dynamics under the background of hydrodynamical evolving bulk matter.

\subsection{Initial production of heavy quarks and fragmentation}
\label{sec2.4}
The initial heavy quarks momentum distribution is obtained up to leading order (LO) with the centrality dependent nuclear parton distribution function EPS09~\cite{EPS09}. The initial transverse positions of heavy quarks are distributed according to the nuclear overlap function of the colliding nuclei in the Glauber model approach. For further details, we refer to the Ref.~\cite{Akamatsu}. With these initial phase space distribution of heavy quarks, we perform the Langevin diffusion of the heavy quarks as described earlier in Sec.\ref{sec2.1}. After Langevin evolution, we hadronize the heavy quarks (charm($c$) and bottom($b$)) via Petersen fragmentation function~\cite{Peterson} to obtain the final momentum distribution of HF mesons ($D$ and $B$). In the Petersen fragmentation function, the parameters we have used are, $\epsilon_{c} = 0.016$ for $c$ quarks to $D$ mesons and $\epsilon_{b} = 0.0012$ for $b$ quarks to $B$ mesons.

Finally, the nuclear modification factor, $R_{AA}$ is defined as,

\begin{equation}
R_{AA} =  \frac{dN^{AA}/dp_{T}}{N_{coll}dN^{pp}/dp_{T}}
\end{equation}
where, $N_{coll}$ is the number of binary nucleon-nucleon collisions for a given centrality class, obtained from Glauber model calculations.

\subsection{Hydrodynamic evolution of the background QGP medium}
\label{sec2.5}
In heavy-ion collisions, the produced hot and dense  QCD medium is in the pre-equilibrium phase before it reaches local thermalization. Generally, the QCD medium undergoes rapid thermalization at time around $\tau_{0} = 0.6$ $fm$ and the hydrodynamic evolution begins.

In this work, the hydrodynamic evolution is understood by ($3+1$)-dimensional relativistic viscous hydrodynamics, vHLLE~\cite{Karpenko}. We assume initial time $\tau_{0} = 0.6$ $fm$, critical temperature $T_{c} = 150$ MeV, shear viscosity $\eta/s = 0.08$
and bulk viscosity $\zeta/s = 0.04$ in the hadronic phase for $Au-Au$ and $Pb-Pb$ collisions. We use optical Glauber initial state for this hydrodynamic evolution. It provides the space-time history of the flow velocity and temperature of the evolving medium. This information of space-time history is used in performing the heavy flavour Langevin simulation. 

\section{Numerical Results and Discussions}
\label{sec3}

\begin{figure} 
	\begin{minipage}{\columnwidth}
		\centering
		\includegraphics[width=\textwidth]{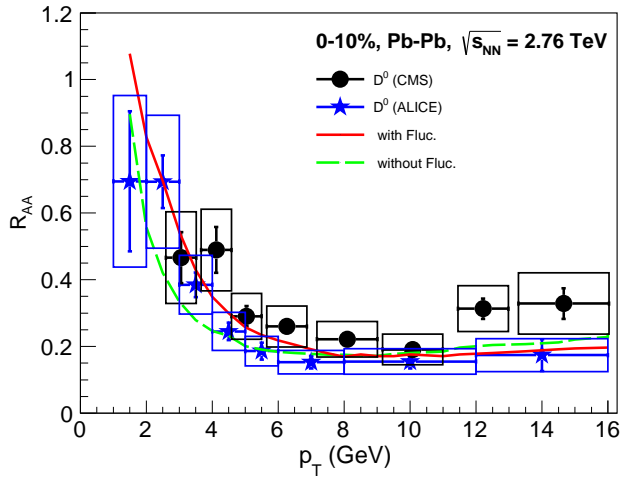}
		\caption{The nuclear modification factor $R_{AA}$ of $D^{0}$ mesons with the effect of fluctuations as a function of transverse momentum $p_{T}$ for $0-10$$\%$ centrality in $Pb-Pb$ collisions at $\sqrt{s_{NN}} = 2.76$ TeV, compared with the ALICE~\cite{ALICE_D} and CMS~\cite{CMS_D_2TeV} data. }
		\label{RaaCent0102TeV}
	\end{minipage}	
\end{figure}

\begin{figure} 
	\begin{minipage}{\columnwidth}
		\centering
		\includegraphics[width=\textwidth]{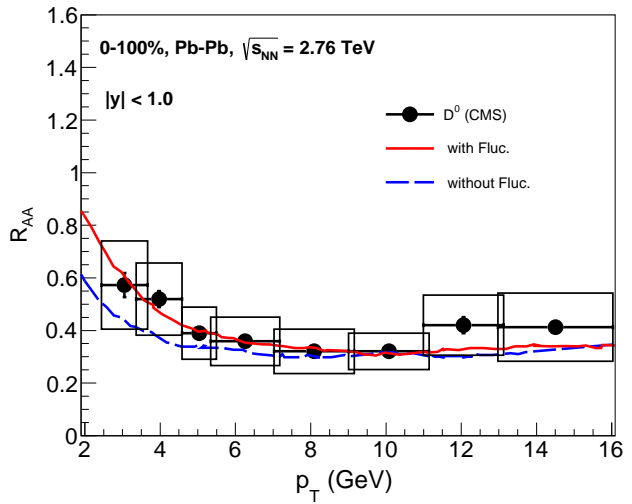}
		\caption{The nuclear modification factor $R_{AA}$ of $D^{0}$ mesons with the effect of fluctuations as a function of transverse momentum $p_{T}$ for $0-100$$\%$ centrality in $Pb-Pb$ collisions at $\sqrt{s_{NN}} = 2.76$ TeV, compared with the CMS~\cite{CMS_D_2TeV} data.}
		\label{RaaCent01002TeV}
	\end{minipage}	
\end{figure}

In Figs.\ref{RaaCent0102TeV} and \ref{RaaCent01002TeV}, we display the nuclear modification factor, $R_{AA}$, of $D^{0}$ mesons with the effect of fluctuations as a function of transverse momentum $p_{T}$ in $Pb-Pb$ collisions at $\sqrt{s_{NN}} = 2.76$ TeV for $0-10$$\%$ and $0-100$$\%$ centralities respectively. The results are compared with the $R_{AA}$ measurements of $D^{0}$ mesons by ALICE~\cite{ALICE_D} and CMS~\cite{CMS_D_2TeV} Collaborations. It is observed that both the results, with and without fluctuations describe the experimental measurements within their systematic uncertainties. However, at the momentum region ($p_{T} < 8$ GeV), the results with the effect of fluctuations describe the experimental data well compared to the case of without fluctuations, i.e., the fluctuations have significant effects on the nuclear modification factor $R_{AA}$ of $D^{0}$ mesons at the lower momentum region. Since, the fluctuations cause energy gain to the charm quarks, significant at the lower momentum limit as discussed in Sec.\ref{sec2.3}.

\begin{figure} 
	\begin{minipage}{\columnwidth}
		\centering
		\includegraphics[width=\textwidth]{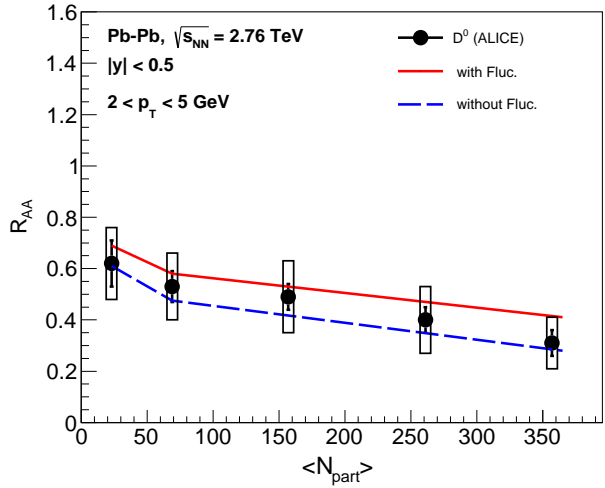}
		\caption{The nuclear modification factor $R_{AA}$ of $D^{0}$ mesons with the effect of fluctuations as a function of $N_{part}$ in $Pb-Pb$ collisions at $\sqrt{s_{NN}} = 2.76$ TeV, compared with the ALICE~\cite{ALICE_DNpart} data.}
		\label{RaaNpart2TeV}
	\end{minipage}	
\end{figure}

Fig.\ref{RaaNpart2TeV} shows the nuclear modification factor, $R_{AA}$, of $D^{0}$ mesons with the effect of fluctuations as a function of $N_{part}$ in $Pb-Pb$ collisions at $\sqrt{s_{NN}} = 2.76$ TeV. We compare the obtained results with the experimental measurements of $R_{AA}$ of $D^{0}$ mesons by ALICE~\cite{ALICE_DNpart} Collaboration. We observe that the results, with and without fluctuations, describe the experimental data within their systematic uncertainties. The effect of fluctuations causes less suppression of $D^{0}$ mesons w.r.t without fluctuations for all values of $N_{part}$, i.e., the fluctuations have a prominent effect on $R_{AA}$ of $D^{0}$ mesons for all centrality classes of $Pb-Pb$ collisions at $\sqrt{s_{NN}} = 2.76$ TeV.

\begin{figure} 
	\begin{minipage}{\columnwidth}
		\centering
		\includegraphics[width=\textwidth]{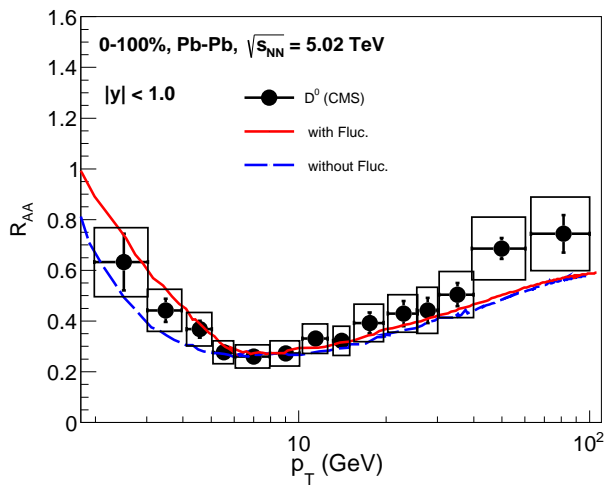}
		\caption{The nuclear modification factor $R_{AA}$ of $D^{0}$ mesons with the effect of fluctuations as a function of transverse momentum $p_{T}$ for $0-100$$\%$ centrality in $Pb-Pb$ collisions at $\sqrt{s_{NN}} = 5.02$ TeV. The experimental data are taken from CMS Collaboration~\cite{CMS_D_5TeV}. }
		\label{RaaD5TeV}
	\end{minipage}	
\end{figure}

\begin{figure} 
	\begin{minipage}{\columnwidth}
		\centering
		\includegraphics[width=\textwidth]{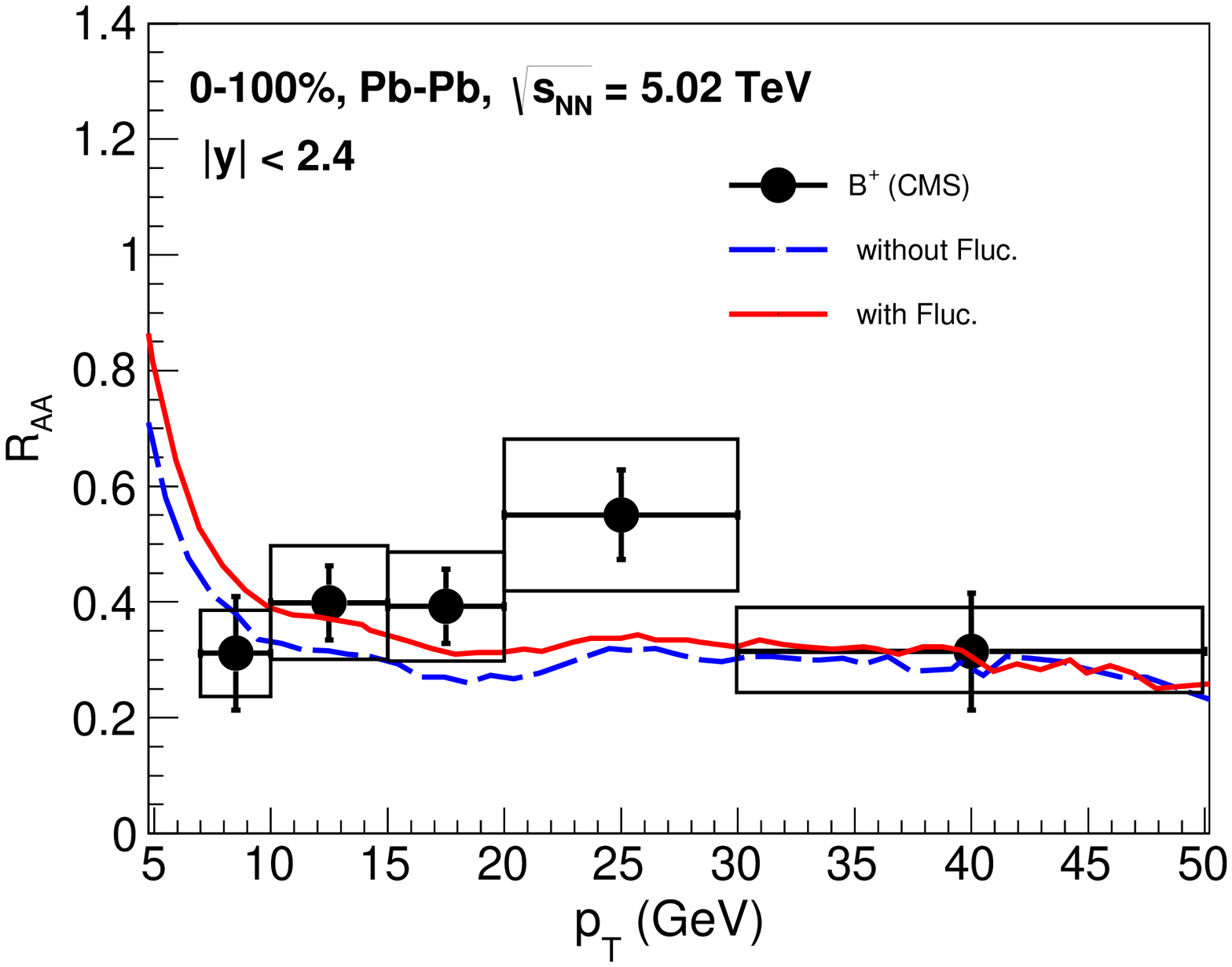}
		\caption{The nuclear modification factor $R_{AA}$ of $B^{+}$ mesons with the effect of fluctuations as a function of transverse momentum $p_{T}$ for $0-100$$\%$ centrality in $Pb-Pb$ collisions at $\sqrt{s_{NN}} = 5.02$ TeV. The experimental data are taken from CMS Collaboration~\cite{CMS_B_5TeV}.}
		\label{RaaB5TeV}
	\end{minipage}	
\end{figure}
%\FloatBarrier
\begin{figure*}[!]
	\centering
	\includegraphics[width=0.9\textwidth]{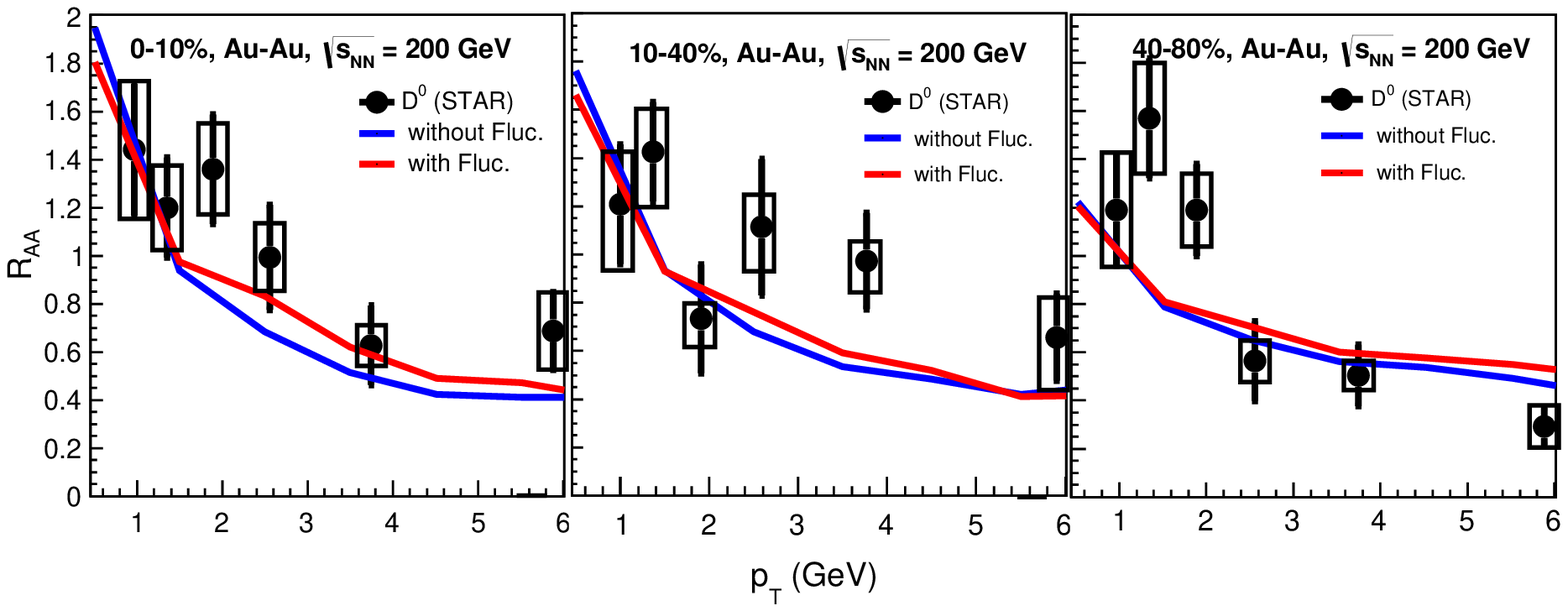}
	\caption{The nuclear modification factor $R_{AA}$ of $D^{0}$ mesons at mid-rapidity with the effect of fluctuations as a function of transverse momentum $p_{T}$ for $0-10$$\%$ (left), $10-40$$\%$ (middle) and $40-80$$\%$ (right) centrality bins in $Au-Au$ collisions at $\sqrt{s_{NN}} = 200$ GeV. The experimental data are taken from STAR Collaboration~\cite{Rhic}.}
\label{RaaRHIC}
\end{figure*}

%\lipsum[4]

Figs.\ref{RaaD5TeV} and \ref{RaaB5TeV} display the nuclear modification factor, $R_{AA}$, with the effect of fluctuations as a function of transverse momentum $p_{T}$ for $0-100$$\%$ centrality in $Pb-Pb$ collisions at $\sqrt{s_{NN}} = 5.02$ TeV for $D^{0}$ and $B^{+}$ mesons respectively. The obtained results are compared with the CMS data~\cite{CMS_D_5TeV,CMS_B_5TeV}. We notice that the results obtained without fluctuations are close to the experimental measurements within their uncertainties for $D^{0}$ and $B^{+}$ mesons as measured by CMS experiment. However, when the effects of the fluctuations are taken into account, the results are in good agreement with the experimental data significantly at the lower momentum region. This is because of the fact that the chromo-electromagnetic field fluctuations cause charm and bottom quarks to gain energy which is substantial in the lower momentum limit and this energy gain is reflected in the nuclear modification factor of $D^{0}$ and $B^{+}$ mesons as measured in the experiments.

In Fig.\ref{RaaRHIC},  we show the nuclear modification factor, $R_{AA}$, of $D^{0}$ mesons as a function of transverse momentum $p_{T}$ for $0-10$$\%$ (left), $10-40$$\%$ (middle) and $40-80$$\%$ (right) centrality classes in $Au-Au$ collisions at $\sqrt{s_{NN}} = 200$ GeV with the effect of field fluctuations included. The obtained results are compared with the $R_{AA}$ measurements of $D^{0}$ mesons by STAR Collaboration~\cite{Rhic}. We observe that the obtained results, both with and without fluctuations, close to the experimental data in all centrality bins. However the results with the effect of fluctuations show less suppression compared to the case of without fluctuations, since the fluctuations cause charm quarks to gain energy.

Here we remark on the possible theoretical uncertainties that may appear in our calculations. The drag coefficients of heavy quarks in QGP medium is not a settled issue. We have estimated the drag coefficients from energy loss calculations of heavy quarks. In that calculations, we have used the fixed coupling constant instead of running coupling. The semiclassical approximation (which is equivalent to the hard thermal loop approximation in the weak coupling limit) has been made to calculate mean energy loss and energy gain of heavy quarks due to field fluctuations. The fragmentation function used in the hadronization involves uncertainties and that uncertainties might affect the $R_{AA}$.

\section{Summary and Conclusions}
\label{sec4}
The heavy quark energy loss and the chromo-electromagnetic field fluctuations in the QGP is important enough for the phenomenology of heavy quark jet propagation and suppression. In this work, we focus to study the effect of the chromo-electromagnetic field fluctuations on heavy quark dynamics and HF suppressions in QGP medium in the ambit of Langevin dynamics. We have taken care of the collisional and radiative energy loss while estimating the drag coefficients. The Langevin dynamics for heavy quarks is solved in the background of evolving QGP medium. We calculate the nuclear modification factor $R_{AA}$ of $D$ mesons in $Au-Au$ collisions at $\sqrt{s_{NN}} = 200$ GeV; $R_{AA}$ of $D$ and $B$ mesons in $Pb-Pb$ collisions at $\sqrt{s_{NN}} = 2.76$ and $5.02$ TeV. We note that the effect of the chromo-electromagnetic field fluctuations is less at RHIC energy. The obtained results without these fluctuations are close to the experimental measurements at RHIC energy. Nevertheless, the data can also be described with the effect of the fluctuations. Besides that, the fluctuations have considerable effects on the suppressions of $D$ and $B$ mesons at lower transverse momentum in LHC energies. The results without fluctuations are close enough to the ALICE and CMS experimental measurements within their uncertainties. But the measurements can be described well at the lower transverse momentum region when the effect of the fluctuations is taken into account. The field fluctuations have an important impact on the heavy quark diffusion in QGP medium and in describing the experimentally observed HF suppressions.

\begin{acknowledgements}
	We are very much grateful to Yuriy Karpenko for providing us the outputs of hydrodynamic evolutions. We also acknowledge Santosh K. Das, Jane Alam, Munshi G. Mustafa, Sandeep Chatterjee and Prithwish Tribedy for fruitful discussions. At last but not least, we thank to the VECC grid computing team for constantly maintaining the grid facility active when the work was done.
\end{acknowledgements}

\end{document}